\pdfoutput=1
\documentclass{llncs}
\usepackage{listings}
\usepackage{tikz}
\usepackage{graphicx}
\usepackage{amsmath}
\usepackage{latexsym}
\usepackage[hidelinks]{hyperref}
\usetikzlibrary{positioning,shapes,arrows,automata}
\usepackage[vlined,ruled,linesnumbered]{algorithm2e}

\newcommand{\nassimscomment}[1]{
\begin{center}
\fbox{
\begin{minipage}{2.5in}
{\bf Nassim's comment:} {\it #1}
\end{minipage}
}
\end{center}
}

\begin{document}

\title{Data-Flow Guided Slicing}

\author{Mohamed Nassim Seghir}
\institute{University College London}

\definecolor{codegreen}{rgb}{0,0.6,0}
\definecolor{codegray}{rgb}{0.5,0.5,0.5}
\definecolor{codepurple}{rgb}{0.58,0,0.82}
\definecolor{backcolour}{rgb}{0.95,0.95,0.92}
\definecolor{pblue}{rgb}{0.13,0.13,1}
\definecolor{pgreen}{rgb}{0,0.5,0}
\definecolor{pred}{rgb}{0.9,0,0}
\definecolor{pgrey}{rgb}{0.46,0.45,0.48}

 \lstdefinestyle{mystyle2}{
  showspaces=false,
  showtabs=false,
  breaklines=true,
  showstringspaces=false,
  breakatwhitespace=true,
  commentstyle=\color{pgreen},
  keywordstyle=\color{pblue},
  stringstyle=\color{pred},
  basicstyle=\ttfamily,
  moredelim=[il][\textcolor{pgrey}]{},
  moredelim=[is][\textcolor{pgrey}]{}{}
}

 \lstdefinestyle{mystyle}{
    backgroundcolor=\color{backcolour},   
    commentstyle=\color{codegreen},
    keywordstyle=\color{magenta}\bfseries,
    numberstyle=\tiny\color{codegray},
    stringstyle=\color{codepurple}
}

\lstset{
language=Java,
basicstyle=\ttfamily\footnotesize,
numbers=left,
morekeywords={assert, assume},
style=mystyle,
frame=none
}

\newtheorem{Remark}{Remark}[section]
\newtheorem{Example}{Example}[section]
\newcommand{\wuntilop}{\;{\underline{\mathcal{WU}}}\;}
\newcommand{\true}{\mathsf{true}}
\newcommand{\false}{\mathsf{false}}
\newcommand{\program}{P}
\newcommand{\tr}{\pi}
\newcommand{\loc}{\ell}
\newcommand{\stmt}{st}

\maketitle

\begin{abstract}
We propose a flow-insensitive analysis that prunes out portions of code which are irrelevant to a specified set of data-flow paths. Our approach is fast and scalable, in addition to being able to generate a certificate as an audit for the computed result. We have implemented our technique in a tool called DSlicer and applied it to a set of 10600 real-world Android applications. Results are conclusive, we found out that the program code can be significantly reduced by 36\% on average with respect to a specified set of data leak paths. 
\end{abstract}

\section{Introduction}
Applying static analysis naively may result in exploring parts of code which are irrelevant to a property of interest.
We propose a lightweight approach for slicing programs that can be invoked by other tools as a preprocessing phase. As slicing criterion, our technique accepts a set of data-flow paths and returns, as result, a reduced program which is free from parts of code irrelevant to the specified set of paths. Hence, the resulting program is guaranteed to include all the specified data-flows. This can benefit a more precise (heavyweight) data-flow analysis allowing it to operate on a reduced version of the original program.  

The effectiveness of our approach is the result of its compact representation, allowing it to efficiently encode programs without sacrificing soundness with respect to the slicing criterion. Moreover, our technique can produce an independently checkable certificate attesting its soundness, hence providing a formal argument for trust.   

\begin{figure}[h]
\begin{center}
\begin{tabular}{@{\hspace{-.1in}}c@{\hspace{0.5in}}c@{\hspace{-0.1in}}} 
\hline \\
\begin{minipage}{6cm}
\begin{tabular}{c}
\begin{lstlisting}
  class A 
  {  
    public static void main()
    {
      C o = new C();
      o.m1();
      o.m3();
      o.m4();
      o.m5();
    }
  }
\end{lstlisting}
\\
{\bf(a)}
\\
\\
\begin{tikzpicture}[->,>=stealth',shorten >=1pt,auto,node distance=1.3cm,semithick]
  \tikzstyle{srcsink}=[rectangle,rounded corners,thick,draw=blue!75,fill=blue!20]
  \tikzstyle{varid}=[]
  
  \node[srcsink]         (SR)  {$SR$};
  \node[varid]         (Cm2v) [below left of=SR] {$C.m2.v$};
  \node[varid]         (Cm2r) [below of=Cm2v] {$C.m2.r$};
  \node[varid]         (Cm1v) [below of=Cm2r] {$C.m1.v$};
  \node[varid]         (Cv1) [below right of=SR] {$C.v1$};
  \node[varid]         (Cv2) [below of=Cv1] {$C.v2$};
  \node[srcsink]         (SK) [below of=Cv2] {$SK$};
  \node[varid]         (0) [below of=SK] {$0$};

  \path (SR) edge node {}  (Cm2v)
  (Cm2v) edge node {} (Cm2r)
  (Cm2r) edge node {} (Cm1v)
  (SR) edge node {}  (Cv1)
  (Cv1) edge node {}  (Cv2)
  (Cv2) edge node {}  (SK)
  (0) edge node {}  (SK)
;
\end{tikzpicture}
\\
{\bf(c)}
\end{tabular}
\end{minipage}
&
\begin{minipage}{5cm}
\begin{tabular}{c}
\begin{lstlisting}
  class C 
  {  
    private int v1, v2;

    public void m1(){
      int v = m2();
      sink(0);
    }   
    public void m2(){
      int v = source();
      return(v);
    }
    public void m3(){
      v1 = source();
    }
    public void m4(){
      v2 = v1;
    }
    public void m5(){
      sink(v2);
    }
  }
\end{lstlisting}
\\
{\bf(b)}
\end{tabular}
\end{minipage}
\\
&
\\
\hline
\end{tabular}
\end{center}
\caption{Example illustrating the translation of a program to an assignment graph}
\label{fig:code_info_leak}
\end{figure}
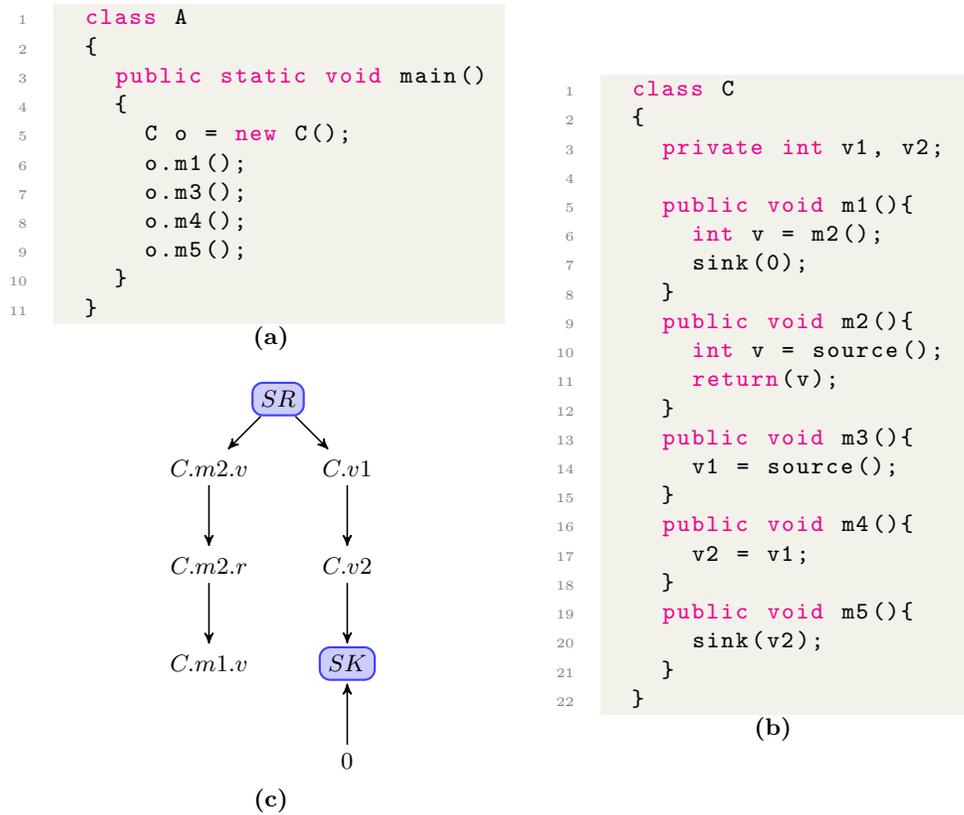

We implemented our approach in a tool called DSlicer and applied it to 10600 real-world Android applications. Results are conclusive, they show that our analysis is efficient (runs in 5 seconds on average), scales up to large programs (containing $>$40000 methods) and dramatically reduces the program size. We found out that the number of methods in the program can be reduced by 36\% on average. This represents a significant saving for potential client tools.
\paragraph{\bf Example.}
To illustrate our approach, let us consider the simple Java example in Figure~\ref{fig:code_info_leak}(a). In the function \textsf{main}, an instance of the class $C$, defined in Figure~\ref{fig:code_info_leak}(b), is created and its methods $m1$, $m3$, $m4$ and $m5$ are consecutively invoked. The function \textsf{source} returns some private (integer) information, it could be for example the identifier of a mobile device. The function \textsf{sink} takes an integer as parameter and sends it out, for instance via the Internet or writes it on the extenal storage. Private information is leaked if there is a path from \textsf{source} to \textsf{sink}. 
We want to identify the methods that are not relevant to information leaks, i.e., no path from \textsf{source} to \textsf{sink} is missed if these methods are removed from the code. This cannot be achieved by simply considering the call graph. For example \textsf{sink} is invoked by $m1$ as well as \textsf{source} which is transitively called through $m2$. However, there is no connection between them, thus $m1$ and $m2$ are not involved in any information leak. On the other hand, $m4$ neither invokes \textsf{source} nor \textsf{sink}, but it is involved in an information leak. It consists of copying the private information stored in $v1$ (generated via $m3$) to $v2$ which is then leaked after calling $m5$. To detect such leaks, our approach proceeds as follows. First, the program is flattened using a flow-insensitive representation that we call \emph{assignment graph}, this is illustrated in Figure~\ref{fig:code_info_leak}(c). Nodes of the graph represent variables in the program  and edges model the information flow (assignments) between them. The procedures \textsf{source} and \textsf{sink} are respectively represented via nodes $SR$ and $SK$. Their implementation is not relevant for the analysis. A local variable $v$ in a method $m$, which is a member of a class $C$, is represented in the graph by $C.m.v$. For example, the assignment in method $m2$ at line 10 is modeled in the graph via $C.m2.v \leftarrow SR$. A field (member) $v$ of a class $C$ is modeled via $C.v$. As an example, the assignment at line 17 is modeled via $C.v2 \leftarrow C.v1$. Hence, for two different objects $O_1$ and $O_2$ belonging to the same class $C$, the field access $O_1.v$ and $O_2.v$ are mapped to the same identifier $C.v$. While this representation is too coarse as the access to the fields of all object instances of the same class in a program are mapped to a unique node in the graph, it has the advantage of soundly approximating aliases when objects are dereferenced. A return statement is modelled via an assignment from the returned variable to a special variable $r$. The return statement at line 11 corresponds to $C.m2.v \rightarrow C.m2.r$ in the graph. Finally, a call assignment to a variable $v$ is represented via an edge from the special variable $r$ of the callee to the return-to variable $v$ of the caller. Line 6 in the code is translated to $C.m2.r \rightarrow C.m1.v$.

To find out which methods are relevant to information leakage paths, it suffices to go through the paths leading from sources $SR$ to sinks $SK$ in the assignment graph and pick up the variables appearing along them. Any method in which those variables appear is considered to be relevant. Back to our example, we only have the path $SR \rightarrow C.v1 \rightarrow C.v2 \rightarrow SK$. The identifiers $C.v1$ and $C.v2$ respectively correspond to the fields $v1$ and $v2$ of class $C$. They appear (are used) in the methods $m3$, $m4$ and $m5$, hence they are the relevant methods. Methods $m1$ and $m2$ can then be safely discarded. In the next section, we cover the translation from a program to an assignment graph in more details. 

\section{Assignment Graph: a Compact Representation}
\begin{table*}[t]
\begin{center}
\begin{tabular}{|c|c|c|c|}
\hline
Instruction & Syntax & Context & Translation\\
\hline
\hline
\textsf{const\_to\_var} & $v := n$ & $v \in m$, $m \in C$ & $C.m.v \leftarrow n$\\
\hline
\hline
\textsf{var\_to\_var} & $v_1 := v_2$ & $v_1, v_2 \in m$, $m \in C$& $C.m.v_1 \leftarrow C.m.v_2$\\
\hline
\hline

\textsf{uni\_op} & $v_1 := \mathsf{op}\;v_2$ & $v_1, v_2 \in m$, $m \in C$&  $C.m.v_1 \leftarrow C.m.v_2$\\
\hline
\hline

\textsf{bin\_op} & $v_1 := v_2 \;\mathsf{op}\;v_3$ & $v_1, v_2, v_3 \in m$, $m \in C$&  $C.m.v_1 \leftarrow C.m.v_2$\\
& & &  $C.m.v_1 \leftarrow C.m.v_3$\\
\hline
\hline

\textsf{var\_to\_array} & $v_1[v_2] := v_3$ & $v_1, v_2, v_3 \in m$, $m \in C$&  $C.m.v_1 \leftarrow C.m.v_3$\\
\hline
\hline

\textsf{array\_to\_var} & $v_1 := v_2[v_3]$ & $v_1, v_2, v_3 \in m$, $m \in C$&  $C.m.v_1 \leftarrow C.m.v_2$\\
\hline
\hline

\textsf{var\_to\_field} & $o.f := v$ & $\mathsf{class}(o) = C$ & $C.f \leftarrow C'.m.v$\\
& & $v \in m$, $m \in C'$ & \\
\hline
\hline

\textsf{field\_to\_var} & $v := o.f$ & $\mathsf{class}(o) = C$ & $ C'.m.v \leftarrow C.f$\\
& & $v \in m$, $m \in C'$ & \\
\hline
\hline

\textsf{call} & $v :=\mathsf{call}\;m',\;[v_0,\ldots,v_n]$ & $v,v_0,\ldots,v_n \in m$& $i \in [0,n]$\\
&& $\textsf{param}(m') = [x_0,\ldots,x_n]$& $C'.m'.x_i \leftarrow C.m.v_i$\\
&& $m \in C, m' \in C'$& $C.m.v \leftarrow C'.m'.r$\\
\hline
\hline

\textsf{return} & $\mathsf{return}\;v$ & $v \in m$, $m \in C$ & $ C.m.r \leftarrow C.m.v$\\
\hline
\end{tabular}
\caption{Instruction syntax and their corresponding representations in the assignment graph.}
\label{fig:assign_inst}
\end{center}
\end{table*}
Our analysis is applied to object oriented programs, therefore we need to account for the main features distinguishing them from simple imperative programs. We use a Jimple-like \cite{Vallee-RaiGHLPS00} intermediary representation which allows to encode most of the language constructs using a minimal set of three-address-based instructions. It has been shown that the whole instruction set of widely deployed virtual machines such as JVM and Dalvik can be encoded in Jimple \cite{Vallee-RaiGHLPS00,BartelKTM12}. 

To ease the presentation, we omit the technical details and consider a core set of the most representative instructions. All the remaining instructions are variants derived from the core set. We do not consider branching instructions as they are irrelevant to our analysis. Table~\ref{fig:assign_inst} illustrates the main instructions considered in our study. Column ``Syntax'' gives the syntax of an instruction and column  ``Translation'' provides its translation in the assignment graph. We have seen in the previous section how instructions \textsf{const\_to\_var}, \textsf{var\_to\_var} are translated. Instructions \textsf{uni\_op} and \textsf{bin\_op} respectively represent unary and binary (arithmetic or logic) operations. Their semantics is irrelevant for us, we are just interested in the information flow they induce. Hence, we have an edge from the operand on the right hand side to the lvalue in case of \textsf{uni\_op} and two edges in case  of \textsf{bin\_op}, modelling the information flow from both operands. Column ``Context'' provides the scope of the different variables, $v \in m$  indicates that $v$ is local (belongs) to method $m$.

In case of an assignment to an array (\textsf{var\_to\_array}), we conservatively assume that the whole array is affected. Keeping track of array indexes requires a non-trivial analysis by itself. We similarly model the opposite case (\textsf{array\_to\_var}).      

For assignments to (from) object fields \textsf{var\_to\_field} (\textsf{field\_to\_var}) having the form $o.f$, we use the class $C$ of the object $o$ as a prefix for the field $f$. As mentioned in the previous section, all dereferences of a given field $f$ of various objects which are instances of the same class $C$ are mapped to a unique node in the graph, namely $C.f$. This has the advantage of soundly approximating aliases. 

For method invocation (\textsf{call}), the receiver object is provided as first argument in case the method is virtual. Hence $v :=\mathsf{call}\;m',\;v_0,\ldots,v_n$ in Table~\ref{fig:assign_inst} is equivalent to $v := v_0.m'(v_1\ldots,v_n)$ ($m'$ is virtual). The translation of such a case consists of adding an edge from each actual parameter of the call to the corresponding formal one. The list of formal parameters of the method $m'$ is obtained via $\textsf{param}$. We also need to assign the returned value to $v$. Thus, we introduce an edge leading from the special variable $r$ of the callee ($m'$) to $v$.  

Finally, for a return statement of the form $\mathsf{return}\;v$, we simply add an edge from $v$ to the special variable $r$ of the related method. 

A key feature of object oriented programs is dynamic dispatch due to polymorphism. The resolution of some method calls can only be done at runtime depending on the type of the receiver object. Taking this into account is crucial for the soundness of our analysis. Hence, we build an over-approximation of the call graph using class-hierarchy-based type estimation of receiver objects \cite{SundaresanHRVLGG00}. 

\section{Slice Computation}
In this section we describe our approach for computing a slice with respect to a set of sources $SR$ and sinks $SK$. The slicing criterion is all data-flow paths $r \leadsto k$ such that $r \in SR$ and $k \in SK$.

This is achieved via algorithm \textsf{ComputeSlice} (Algorithm \ref{alg:ComputeSlice}) which takes as parameters an assignment graph $G$, a set of sources $SR$ and a set of sinks $SK$. Sources and sinks are in general API functions. Strictly speaking, $SR$ is the set of identifiers appearing in the program as parameters or return variables of a source API. Similarly, $SK$ is the set of identifiers appearing in the program as parameters of a sink API. 

The initial step of Algorithm \textsf{ComputeSlice} consists of marking all the source identifiers (nodes) in the graph $G$ with $+$, as illustrated at line 4. It then performs a forward propagation of the marking to the successors of the already marked nodes (lines 5-9). Next, sinks are marked with the symbol $-$ (lines 11-12). The marking is then propagated backward, but the propagation is only restricted to predecessor nodes which have been already marked with $+$ in the forward phase (lines 13-17). The relevant identifiers are the ones marked with both $+$ and $-$ (line 18). Intuitively speaking, these are the nodes reachable from sources and reaching sinks. In our implementation, we offer the option for retrieving an exclusively forward or backward slice.

Our slicing is coarse grained, and conservative as we keep an entire method if at least one of the symbols used in it is relevant. Some methods are affected through their local variables (including parameters), see line 19. Other methods are affected through object field access. In the latter case, if an identifier of the form $c.f$ is relevant, it implies that any access to the field $f$ of some object $o$ which is instantiated from $c$ is relevant. Consequently, any method in which $o.f$ appears is considered as relevant (line 20). Finally the set of relevant methods is returned.
\begin{algorithm}[h!]
\KwIn {assignment graph $G$, source set $SR$, sink set $SK$} 
\KwOut {set of procedure identifiers}
\SetKw{Var}{Var}
\Var set $L$\;

$L$ := $\{id \;|\; id \in SR\}$\;

\ForEach{$id \in L$}
  {
     $\mathsf{mark}(id,G,+)$\;
  }

\While{$L \neq \emptyset$}
{
  select and remove an identifier $id$ from $L$\;
  
  \ForEach{$(id,id') \in G$ s.t. $\neg \mathsf{marked}(id',G,+)$}
  {
    $\mathsf{mark}(id',G,+)$\;
    $L$ := $L \cup \{id'\}$\;
  }
 }

$L$ := $\{id \;|\; id \in SK\}$\;

\ForEach{$id \in L$}
  {
     $\mathsf{mark}(id,G,-)$\;
  }

\While{$L \neq \emptyset$}
{
  pick up an identifier $id$ from $L$\;
  
  \ForEach{$(id',id) \in G$ s.t. $\mathsf{marked}(id',G,+) \wedge \neg \mathsf{marked}(id',G,-)$}
  {
    $\mathsf{mark}(id',G,-)$\;
    $L$ := $L \cup \{id'\}$
  }
 }

$ids$ := $\{id \;|\; id \in \mathsf{nodes}(G) \wedge \mathsf{marked}(id,G,+) \wedge \mathsf{marked}(id,G,-)\}$\;

$via\_locals$ := $\{m \;|\; \exists c \;\exists v\;\; c.m.v \in ids\}$\;

$via\_fields$ :=  $\{m \;|\; \exists o \; \exists f\;\; o.f \text{ used in }m \wedge \mathsf{class}(o) = c \wedge c.f \in ids   \}$\; 

$relevant$ := $via\_locals \cup via\_fields$\;

\Return{$relevant$} 
\caption{ComputeSlice}
\label{alg:ComputeSlice}
\end{algorithm}

\paragraph{\bf Certification.}
A key question that may arise is: how can we trust the result (soundness) of the analysis? For this we return a certificate which can be independently checked to confirm or refute the result of the analysis. The certificate is made of two parts:
\begin{itemize}
\item {\bf  Translation certificate}: it consists of the assignment graph. It allows to check that the program is faithfully translated according to the rules presented in Table \ref{fig:assign_inst}. The checking is quite straightforward, it suffices to iterate over the instructions of each method and check that their corresponding translations are present in the graph.
\item {\bf Analysis certificate}: consists of a mapping assigning each node of the graph to the marking ($+$, $-$) generated during the slice computation. Checking the certificate in this case simply consists of verifying that every source is marked with $+$, every sink is marked with $-$, and for any edge $n_1 \rightarrow n_2$, check that $\textsf{marked}(n_1,+) \Rightarrow \textsf{marked}(n_2,+)$ and $\textsf{marked}(n_2,-) \Rightarrow \textsf{marked}(n_1,-)$. This is carried out in a single linear pass as opposed to the slice computation which requires a forward and backward pass. We provide an implementation of a certificate checker in our tool.
\end{itemize}

\section{DSlicer: Implementation and Experiments} 

We implemented our approach in a tool called DSlicer, which is written in Python and uses Androguard\footnote{https://github.com/androguard} as front-end for parsing and decompiling Android applications. It takes as input an Android application in bytecode format (APK) and a configuration file containing the sources and sinks to be considered in the analysis. One can tailor the analysis by modifying the configuration file. As output, DSlicer returns a list of methods that are relevant (irrelevant) to the slicing criterion, (optionally) with an accompanying certificate. It also has an option for generating a new application (APK) that is free from irrelevant methods. In checking mode, DSlicer accepts an application and a certificate and answers whether the certificate is valid with respect to the application taken as input.   

\paragraph{\bf Experiments.} 
We performed experiments on a large set of 10600 Android applications collected from various sources, including the Google Play store. As slicing criterion, we used the set of sources and sinks provided with Flowdroid \cite{Flowdroid}. Therefore, the considered criterion is potential data leak paths.

\begin{figure*}
   \centering
   \begin{tabular}{@{}c@{\hspace{.1cm}}c@{}}
       \includegraphics[width=.50\textwidth]{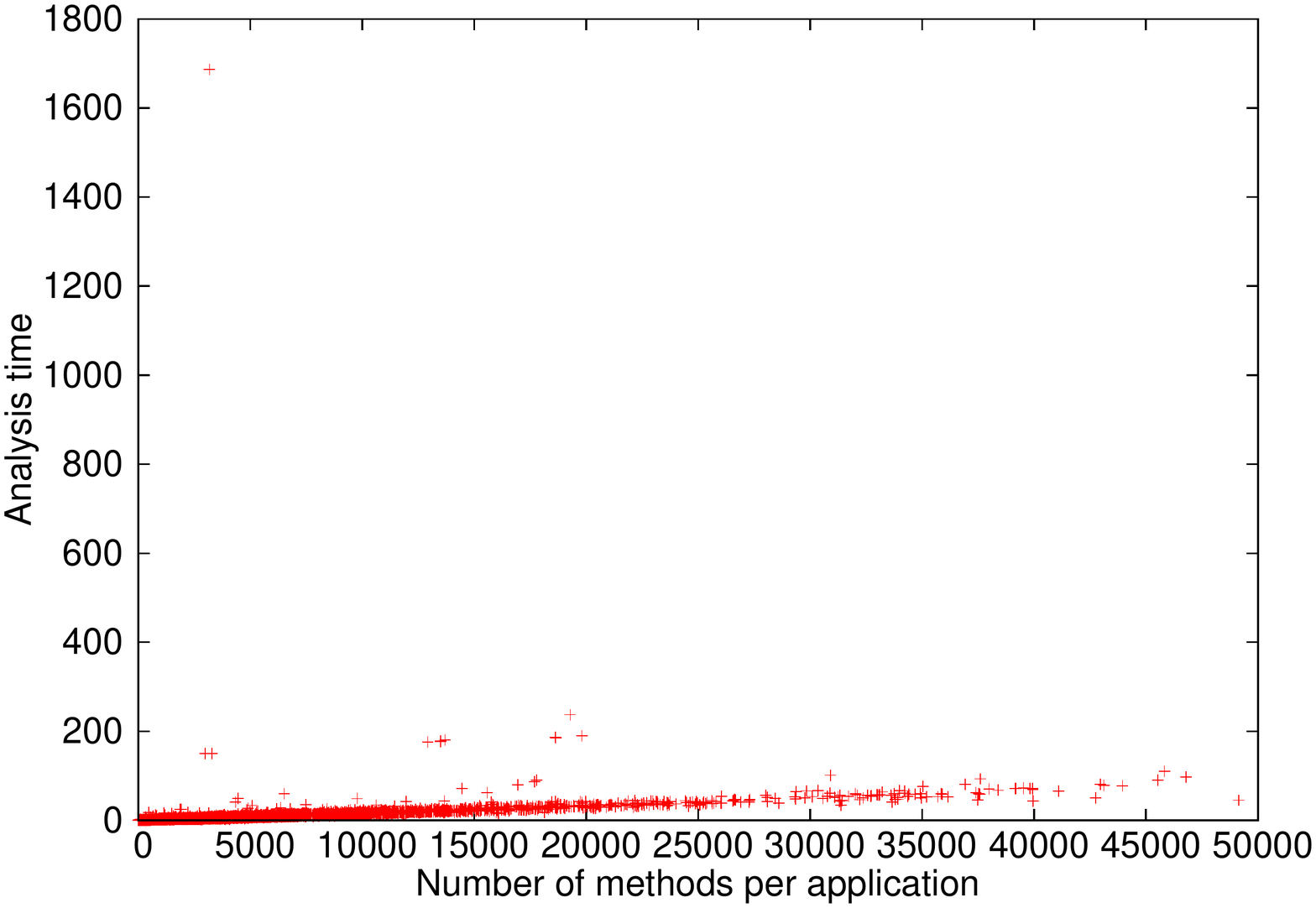} & 
       \includegraphics[width=.50\textwidth]{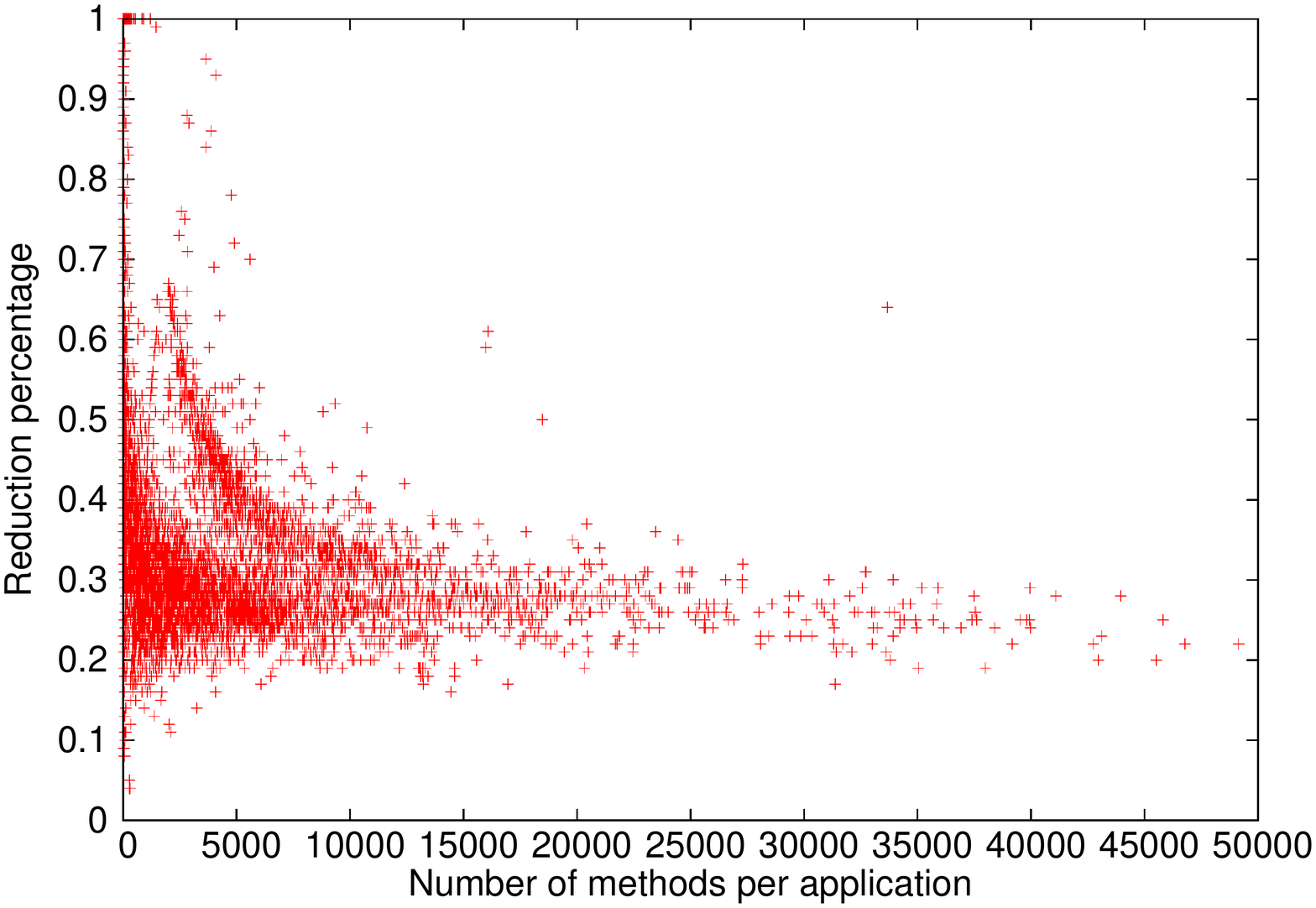} 
       \\
       (a) & (b)
       
   \end{tabular}
 \caption{Time taken by the analysis per application (a) and application size reduction percentage (b) in function of method number}
 \label{fig:graphs}
\end{figure*}

Figure \ref{fig:graphs} illustrates the results in terms of analysis run-time (a) and size reduction percentage (b) per application. The first diagram (a) clearly shows that the run-time of the analysis is linear in the application size (number of methods). Moreover, most of the applications are analyzed in less than 200 seconds. The diagram on the right-hand side (b) represents the gain in application size (irrelevant methods) in function of the method number. We can see that the percentage of methods that can be discarded is between 15\% and 65\% for most of the applications and tends to stabilize between 30\% and 40\% when the number of methods increases. Applications are of decent size (3914 methods on average) with some of them containing up to 49146 methods. The overall running time of DSlicer is 1682 seconds in the worst case, but runs in 5 seconds on average. This is not a significant overhead for potential clients which are far more time consuming. Finally, the most important result is the potential reduction of code size thanks to the present analysis. We found out that 36\% of the methods are in general irrelevant to the considered set of sources and sinks. Hence, they can be safely discarded. Such an important reduction can significantly benefit client tools.  
\section{Related work} 
Our work is at the intersection of several topics:  program slicing, data-flow analysis, and program certification.

Program slicing was originally proposed by Weiser \cite{Weiser81}. While the initial proposal defined a slice as a set of statements that might influence a program point of interest (criterion), a dynamic variant was examined as well \cite{AgrawalH90}. 

Slicing has been applied for various purposes: program debugging \cite{AgrawalDS93}, testing \cite{HarmanHLMW07} comprehension \cite{KorelR98}, re-use \cite{CanforaLM98} and re-engineering \cite{RepsR95}. There are also  semantics-preserving variants \cite{HarmanD97}, allowing changes to the program syntax as long as the program semantics is preserved. In the context of software model checking, path slicing was also proposed \cite{JhalaM05}. 
Rasthofer et al applied slicing for extracting runtime values to increase the recall of existing static analysis \cite{SiegSME16}.
While previous work requires program-semantics to be preserved, our requirement is more relaxed. We only need data-paths to be preserved.

The assignment-graph-based representation used in our approach is inspired by the work of Sundaresan et al \cite{SundaresanHRVLGG00} on the estimation of object dynamic types. Our application makes a more general usage of that representation as we consider more cases for flows such as arithmetic and logic expressions.

Flow-insensitive analyses have been well studied in the literature \cite{ShapiroH97,FosterFA00,HardekopfL09,BlackshearCSS11,AdamsBDLRSW02}. All these approaches are related to pointer analysis. The technique presented by Adam et al \cite{AdamsBDLRSW02} is the closest to our work. They run a scalable, control-flow-insensitive pointer analysis to obtain a conservative over-approximation of the dataflow facts and the statements that must be processed by a client analysis. In addition to being specific to pointer analysis, their approach is targeting C programs. Our method deals with object oriented programs which have more complicated features such as dynamic dispatching, etc.   

There are many data-flow analysis tools for finding information leaks: Flowdroid \cite{Flowdroid}, Taintdroid \cite{Enck:2014}, Amandroid \cite{WeiROR14}, Andromeda \cite{TrippPCCG13} and Droidsafe \cite{GordonKPGNR15}. They represent potential clients for DSlicer as they perform a quite precise analysis. Reducing the size of the code to be analysed can boost their performance.

\section{Conclusion and Further Work}  
We presented a lightweight slicing approach for object oriented programs that can detect and discard parts of code that are irrelevant to a specified set of data-flow paths. A distinguishing feature of our approach lies in the compact representation it uses, allowing it to efficiently encode programs without sacrificing soundness with respect to the slicing criterion. Moreover, our analysis generates an independently checkable certificate attesting the soundness of the returned result. We have implemented our approach in a tool called DSlicer and applied it to 10600 real-world applications. Results show that our approach is effective as it runs in 5 seconds for most of the programs, scales up to large programs ($>$ 40000 methods) and significantly reduces application code size (number of methods) by up to 36\% on average. This is a pertinent optimization that could significantly benefit potential client tools.  
\bibliographystyle{abbrv} 
\bibliography{biblio}

\begin{thebibliography}{10}

\bibitem{AdamsBDLRSW02}
S.~Adams, T.~Ball, M.~Das, S.~Lerner, S.~K. Rajamani, M.~Seigle, and W.~Weimer.
\newblock Speeding up dataflow analysis using flow-insensitive pointer
  analysis.
\newblock In {\em {SAS}}, pages 230--246, 2002.

\bibitem{AgrawalDS93}
H.~Agrawal, R.~A. DeMillo, and E.~H. Spafford.
\newblock Debugging with dynamic slicing and backtracking.
\newblock {\em Softw., Pract. Exper.}, 23(6):589--616, 1993.

\bibitem{AgrawalH90}
H.~Agrawal and J.~R. Horgan.
\newblock Dynamic program slicing.
\newblock In {\em Proceedings of the {ACM} SIGPLAN'90 Conference on Programming
  Language Design and Implementation (PLDI), White Plains, New York, USA, June
  20-22, 1990}, pages 246--256, 1990.

\bibitem{Flowdroid}
S.~Arzt, S.~Rasthofer, C.~Fritz, E.~Bodden, A.~Bartel, J.~Klein, Y.~L. Traon,
  D.~Octeau, and P.~McDaniel.
\newblock Flowdroid: precise context, flow, field, object-sensitive and
  lifecycle-aware taint analysis for android apps.
\newblock In {\em PLDI}, page~29, 2014.

\bibitem{BartelKTM12}
A.~Bartel, J.~Klein, Y.~L. Traon, and M.~Monperrus.
\newblock Dexpler: converting android dalvik bytecode to jimple for static
  analysis with soot.
\newblock In {\em {SOAP}}, pages 27--38, 2012.

\bibitem{BlackshearCSS11}
S.~Blackshear, B.~E. Chang, S.~Sankaranarayanan, and M.~Sridharan.
\newblock The flow-insensitive precision of andersen's analysis in practice.
\newblock In {\em {SAS}}, pages 60--76, 2011.

\bibitem{CanforaLM98}
G.~Canfora, A.~D. Lucia, and M.~Munro.
\newblock An integrated environment for reuse reengineering {C} code.
\newblock {\em Journal of Systems and Software}, 42(2):153--164, 1998.

\bibitem{Enck:2014}
W.~Enck, P.~Gilbert, B.-G. Chun, L.~P. Cox, J.~Jung, P.~McDaniel, and A.~N.
  Sheth.
\newblock Taintdroid: An information flow tracking system for real-time privacy
  monitoring on smartphones.
\newblock {\em Commun. ACM}, 57(3):99--106, Mar. 2014.

\bibitem{FosterFA00}
J.~S. Foster, M.~F{\"{a}}hndrich, and A.~Aiken.
\newblock Polymorphic versus monomorphic flow-insensitive points-to analysis
  for {C}.
\newblock In {\em {SAS}}, pages 175--198, 2000.

\bibitem{GordonKPGNR15}
M.~I. Gordon, D.~Kim, J.~H. Perkins, L.~Gilham, N.~Nguyen, and M.~C. Rinard.
\newblock Information flow analysis of android applications in droidsafe.
\newblock In {\em {NDSS}}, 2015.

\bibitem{HardekopfL09}
B.~Hardekopf and C.~Lin.
\newblock Semi-sparse flow-sensitive pointer analysis.
\newblock In {\em {POPL}}, pages 226--238, 2009.

\bibitem{HarmanD97}
M.~Harman and S.~Danicic.
\newblock Amorphous program slicing.
\newblock In {\em 5th International Workshop on Program Comprehension {(WPC}
  '97), May 28-30, 1997 - Dearborn, MI, {USA}}, pages 70--79, 1997.

\bibitem{HarmanHLMW07}
M.~Harman, Y.~Hassoun, K.~Lakhotia, P.~McMinn, and J.~Wegener.
\newblock The impact of input domain reduction on search-based test data
  generation.
\newblock In {\em Proceedings of the 6th joint meeting of the European Software
  Engineering Conference and the {ACM} {SIGSOFT} International Symposium on
  Foundations of Software Engineering, 2007, Dubrovnik, Croatia, September 3-7,
  2007}, pages 155--164, 2007.

\bibitem{JhalaM05}
R.~Jhala and R.~Majumdar.
\newblock Path slicing.
\newblock In {\em {PLDI}}, pages 38--47, 2005.

\bibitem{KorelR98}
B.~Korel and J.~Rilling.
\newblock Program slicing in understanding of large programs.
\newblock In {\em 6th International Workshop on Program Comprehension {(IWPC}
  '98), June 24-26, 1998, Ischia, Italy}, pages 145--152, 1998.

\bibitem{SiegSME16}
S.~Rasthofer, S.~Arzt, M.~Miltenberger, and E.~Bodden.
\newblock Harvesting runtime values in android applications that feature
  anti-analysis techniques.
\newblock In {\em {NDSS}}, 2016 (To appear).

\bibitem{RepsR95}
T.~W. Reps and G.~Rosay.
\newblock Precise interprocedural chopping.
\newblock In {\em {SIGSOFT} '95, Proceedings of the Third {ACM} {SIGSOFT}
  Symposium on Foundations of Software Engineering, Washington, DC, USA,
  October 10-13, 1995}, pages 41--52, 1995.

\bibitem{ShapiroH97}
M.~Shapiro and S.~Horwitz.
\newblock Fast and accurate flow-insensitive points-to analysis.
\newblock In {\em {POPL'97}}, pages 1--14, 1997.

\bibitem{SundaresanHRVLGG00}
V.~Sundaresan, L.~J. Hendren, C.~Razafimahefa, R.~Vall{\'{e}}e{-}Rai, P.~Lam,
  E.~Gagnon, and C.~Godin.
\newblock Practical virtual method call resolution for java.
\newblock In {\em OOPSLA}, pages 264--280, 2000.

\bibitem{TrippPCCG13}
O.~Tripp, M.~Pistoia, P.~Cousot, R.~Cousot, and S.~Guarnieri.
\newblock Andromeda: Accurate and scalable security analysis of web
  applications.
\newblock In {\em {FASE}}, pages 210--225, 2013.

\bibitem{Vallee-RaiGHLPS00}
R.~Vall{\'e}e-Rai, E.~Gagnon, L.~J. Hendren, P.~Lam, P.~Pominville, and
  V.~Sundaresan.
\newblock Optimizing java bytecode using the soot framework: Is it feasible?
\newblock In {\em CC}, pages 18--34, 2000.

\bibitem{WeiROR14}
F.~Wei, S.~Roy, X.~Ou, and Robby.
\newblock Amandroid: {A} precise and general inter-component data flow analysis
  framework for security vetting of android apps.
\newblock In {\em {CCS}}, pages 1329--1341, 2014.

\bibitem{Weiser81}
M.~Weiser.
\newblock Program slicing.
\newblock In {\em Proceedings of the 5th International Conference on Software
  Engineering, San Diego, California, USA, March 9-12, 1981.}, pages 439--449,
  1981.

\end{thebibliography}

\end{document}